\documentclass{aa} 

\usepackage[colorlinks]{hyperref}
\hypersetup{colorlinks, linkcolor=blue, citecolor=blue }

\hypersetup{
        colorlinks=true,
        allcolors=blue,
        }
\usepackage{natbib}
\usepackage[varg]{txfonts}
\usepackage{graphicx} 
\begin{document}

\title{Irradiation-driven mass transfer for massive companion stars in supersoft X-rays sources}

\titlerunning{Irradiation-driven mass transfer}
\authorrunning{W. Zhao et al}

\author{Weitao Zhao \inst{1}
\and Xiangcun Meng\inst{2,3,4}
\and Yingzhen Cui \inst{5}}
\institute{Institute of Electromagnetic Wave, Henan normal university, Xinxiang 453007, China\\e-mail: zhaoweitao@htu.edu.cn
\and Yunnan Observatories, Chinese Academy of Sciences, Kunming 650216, China\\e-mail: xiangcunmeng@ynao.ac.cn
\and Key Laboratory for the Structure and Evolution of Celestial Objects, Chinese Academy of Sciences, Kunming 650216, China
\and International Centre of Supernovae, Yunnan Key Laboratory, Kunming 650216, P. R. China
\and Key Laboratory of Optical Astronomy, National Astronomical Observatories, Chinese Academy of Sciences, Beijing 100101, China}

\abstract
 {Supersoft X-ray sources (SSSs) have been proposed as one of the progenitors for Type Ia supernovae. However, the exact origin of the quasi-periodic variability in the optical light curve remains a mystery. }
 {In this work, our goal is to investigate the effect of the feedback of an evolved main-sequence companion star on X-ray irradiation and find whether periodic X-ray irradiation of the companion star could reproduce periodic mass transfer.}
 {Using the Modules for Experiments in Stellar Astrophysics (MESA) code, we modeled the evolutionary track of the companion star under the influence of supersoft X-ray irradiation, and we calculated the resulting mass transfer rate.}
 {We find that the supersoft X-ray heating of the companion star can result in the expansion of the companion, causing it to greatly overflow its Roche lobe and thereby increasing the mass transfer rate. The periodic X-ray irradiation on the companion stars leads to periodic changes in the mass transfer rate. For a given companion star, higher irradiation efficiencies result in a higher mass transfer rate. Additionally, the mass transfer rate increases as the mass of the companion star decreases for a given irradiation efficiency.} 
 {The companion star undergoing thermal timescale mass transfer is periodically irradiated by the X-rays from the WD, which can lead to periodic enhancement of the mass transfer rate. The mechanism could be the origin of the quasi-periodic optical light curve in supersoft X-ray sources.}
 

\maketitle

\section{Introduction}
Type Ia supernovae (SNe Ia) are commonly used as excellent standard candles to measure distances due to their uniform luminosity \citep[]{Perlmutter1999, Riess1998, Howell2011}. SNe Ia are used as cosmological probes to examine the equation of state of dark energy and its evolution with time \citep[]{Howell2011}. However, some basic aspects of SNe Ia remain unclear, such as the progenitor star model and their explosion mechanism \citep[]{Branch1995, Hillebrandt2000, Maoz2014}. The single-degenerate (SD) model is one of the proposed SNe Ia progenitor models. In the SD model, a white dwarf (WD) accretes hydrogen-rich material from a non-degenerate companion star. The accreted material undergoes nuclear burning on the surface of the WD, gradually increasing its mass until the WD reaches the Chandrasekhar mass limit, triggering an SNe Ia explosion \citep[]{Whelan1973, Nomoto1984, Meng2009a, Meng2010, Liu2018}. 
 
 Supersoft X-ray sources (SSSs) are characterized by their soft X-ray spectrum, with temperatures of about $15 - 80 \ \rm eV$, and their X-ray luminosities are very high, ranging from $L_{\rm x} = 10^{36} - 10^{38} \ {\rm erg/s}$ \citep{Greiner1991, Alcock1996, vanTeeseling1996, Kahabka1997, Gansicke1998, Cowley2002}. It is thought that SSS systems involve a WD that accretes material from companion stars at a very high rate of $\dot{M}_{\rm acc} \sim 1-4 \times 10^{-7}\,{\rm M}_{\odot}\,{\rm yr}^{-1}$, allowing stable hydrogen shell burning on the surface layers of the WD \citep[]{vandenHeuvel1992, Nomoto2007}. Thus, SSSs are believed to be the most plausible progenitor models for SNe Ia within the SD model \citep[]{Nomoto1984, vandenHeuvel1992}. One of the most obvious observational features of an SSS is the quasi-periodic transition between the optical high and low states, which is believed to be caused by periodic accretion onto the WD \citep{vandenHeuvel1992, PH.2010}. However, the origin of periodic accretion remains a mystery \citep{Pakull1993,Reinsch1996,Reinsch2000,Southwell1996,Meyer-Hofmeister1997}.
 
 In low-mass X-ray binaries, the low-mass companion star expands toward a new thermal equilibrium state due to X-ray irradiation from the accreting neutron star (NS), providing a mechanism to increase the mass transfer rate \citep{PH.1991, Hameury1993, Harpaz1994, Hameury1997, Ritter2000}. \citet{King1995} investigated the feedback of the irradiated secondary star in cataclysmic variables (CVs) and found that this mechanism could yield a very high mass transfer rate ($10^{-8} - 10^{-6} \, {\rm M}_{\odot}\,{\rm yr}^{-1}$) closing or exceeding the stable hydrogen burning rate of the WD \citep{Ritter2000, Harpaz1991, Harpaz1994, Ginzburg2021}.
 It is worth noting that SSSs exhibit a very high X-ray luminosity (${10^{36} - 10^{38} \ { \rm erg/s}}$) and short orbital periods (a few hours to a few days) \citep{Greiner1991, vanTeeseling1996, Kahabka1997}. This may result in a significant X-ray irradiation effect. However, in the past, the impact of X-ray irradiation on the companion star in SSSs has been overlooked.

 \begin{figure}
        \centering
        \includegraphics[width=0.9\columnwidth]{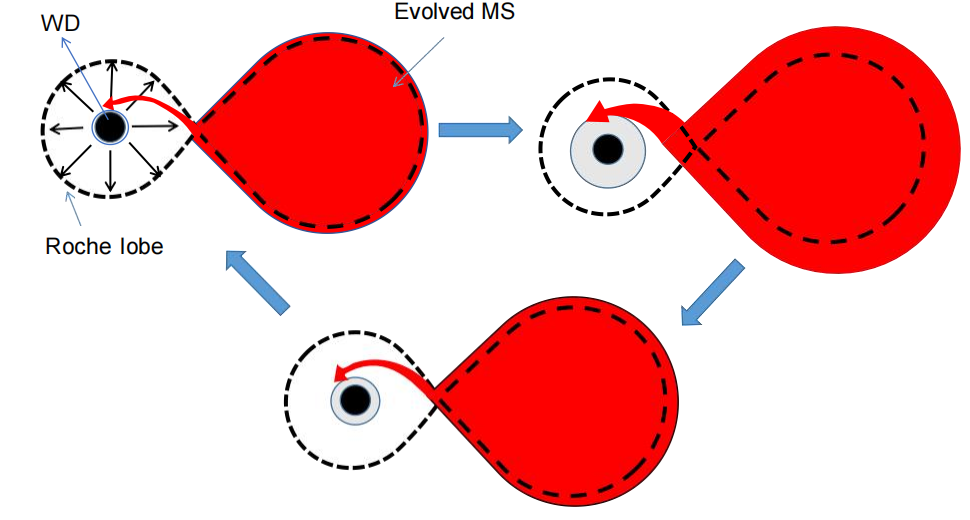}
        \caption{Schematic diagram illustrating the irradiation-driven mass transfer in our model. The black solid circle represents a WD, the blue circle represents the WD photosphere, a red solid circle represents an MS companion star with its Roche lobe overflowed, arrows denote supersoft X-rays, and dashed lines indicate the Roche lobe. }
        \label{fig:explain} 
   \end{figure}
   
 \citet[]{Zhao2022} hypothesize that the companion star is irradiated by periodic supersoft X-rays emitted by the WD, resulting in a periodic mass transfer rate, as illustrated in Fig.~\ref{fig:explain}. Initially, the WD undergoes stable hydrogen burning, emitting supersoft X-rays from its surface. These X-rays irradiate the companion star, prompting it to expand and overflow its Roche lobe. Consequently, the mass transfer rate increases, surpassing the critical accretion rate of the WD and leading to rapid expansion of the WD photosphere. As a result, the surface temperature of the WD decreases, causing a decline in X-ray flux. The companion star gradually contracts, and the mass transfer rate decreases, inducing the contraction of the WD photosphere until it returns to the initial state. Subsequently, the WD emits a significant amount of X-rays again, initiating a cycle that repeats itself. At first, \citet{Zhao2022} successfully reproduced the quasi-periodic optical light curve observed in SSS RX J0513.9-6951 by using a periodic WD mass accretion. However, the existence of periodic mass transfer in an SSS system remains uncertain. 

  In this work, we investigate an SSS system where the primary source is a WD and the companion star is a slightly evolved main-sequence (MS) star that has begun to expand while its central hydrogen is not yet exhausted. The irradiation-driven mass transfer cycles occur when the mass transfer rate between the binary stars is within the stable accretion zone of the WD. In this work, we reproduce, from a modeling perspective, the cycle of irradiation-driven mass transfer. 
  The companion star is irradiated and heated by supersoft X-rays, and we calculated the increased mass transfer rate. In Section~\ref{Method}, we provide a detailed description of our method, and this is followed by the presentation of our results in Section~\ref{Result}. In Section~\ref{discussion}, we discuss our findings, and we summarize our main conclusions in Section~\ref{summary}.

\section{Method}
\label{Method}

  In our model, mass transfer is primarily driven by thermal timescale mass transfer mechanisms. The initial model is a binary system where the primary source is a WD and the companion is an MS star. The companion star, due to its own nuclear evolution, gradually expands to overflow its Roche lobe, resulting in mass transfer occurring at a rate reaching the WD stable accretion rate and continuing for a long time. Consequently, the WD triggers stable hydrogen burning and then emits supersoft X-rays. At this point, we investigated the impact of X-ray irradiation on the companion star under these conditions.
  
  In this work, we focus on SSS RX J0513.9-6951. To simulate the evolution of the companion star irradiated by the periodic supersoft X-rays from the WD, we used the Modules for Experiments in Stellar Astrophysics (MESA) code \citep[]{Paxton2011, Paxton2013, Paxton2015, Paxton2018, Paxton2019}. The WD was treated as a point mass, and the mixing length parameter was set to two. The binary orbital period was set to be 0.7628 d \citep[]{Alcock1996, Reinsch1996}.
 
 The companion star in the SSS system is irradiated by supersoft X-rays from the WD. The irradiated energy $L_{\rm irr}$ by the companion star depends on the cross sections of its companion \citep{Kovetz1988},
 \begin{equation}
  L_\mathrm{irr} = \eta \frac{\omega}{4 \pi} L_{\rm x},
 \end{equation}
 where $L_{\rm x}$ is the X-ray luminosity from the accreting WD, which is stably hydrogen burning. The term ${\eta}$ ($< 1$) is the irradiation efficiency between the WDs and the companion stars, representing the fraction of the X-ray luminosity from the WD that can reach the MS companion star due to the shadowing effect of the accretion disk \citep[]{Milgrom1978, Mason1989, Ritter2000}. The term $\omega$ is the solid angle given by
\begin{equation}
\frac{\omega}{4 \pi} = \frac{(1 - \mathrm{sin} \theta)}{2},
\end{equation}
with
\begin{equation}
\theta = \mathrm{cos}^{-1}\left(\frac{R_{2}}{A}\right),
\end{equation}
where $R_{2}$ is the radius of the companion star and $\mathrm{A}$ is the binary separation \citep[]{Kovetz1988, Ritter2000}. 

 We simulated the feedback of the companion star to the supersoft X-ray irradiation under the simplifying assumption of spherical symmetry. For SSS RX J0513.9-6951, we assumed the WD had a mass of $1.3\,{\rm M}_{\odot}$ and the companion star had a metallicity $Z$ of 0.004 \citep{Zhao2022}. In this work, the supersoft X-rays cannot penetrate the companion star entirely. Due to the increasing opacity with depth in the interior of the companion star, the X-ray energy gradually weakens. 
 The penetration depth is calculated by the radiative transfer equation, namely, $I/I_{\rm 0} = e^{-\tau}$. Here, $\tau$ and $I$ are the optical depth and radiation intensity incident upon the surface of the companion star, respectively. Furthermore, the heated layer gradually penetrates deeper through diffusion into the interior of the companion star. 

 As the companion star is irradiated by the X-rays, the radiation energy penetrates the surface of the companion star \citep{Ritter2000, Harpaz1994}. Due to the increase in internal energy of the companion star, the surface layer of the companion star expands, and it greatly overflows its Roche lobe. The materials overflowing the Roche lobe are then transferred to the WD. We calculated the mass transfer rate by using the model proposed by \citet{Kolb1990}, and a detailed derivation can be found in the appendix of \citet{Kolb1990}. In this work, we create some companion star models (masses ranging from 1.7 to 3.0 ${\rm M}_{\odot}$).
 These companion stars have evolved to the point where they are beginning to expand, with their radii having exceeded their Roche radii, and they are continuously losing material at a rate of $3.0 \times 10^{-7}\, {\rm M}_{\odot}\,{\rm yr}^{-1}$. 
 We investigated the effects on the mass transfer rate of varying the companion stars' masses.
 
\section{Results}
\label{Result}

\subsection{Thermal timescale mass transfer}
\label{Irradiation}

 \begin{figure}
        \centering
        \includegraphics[width=1.0\columnwidth]{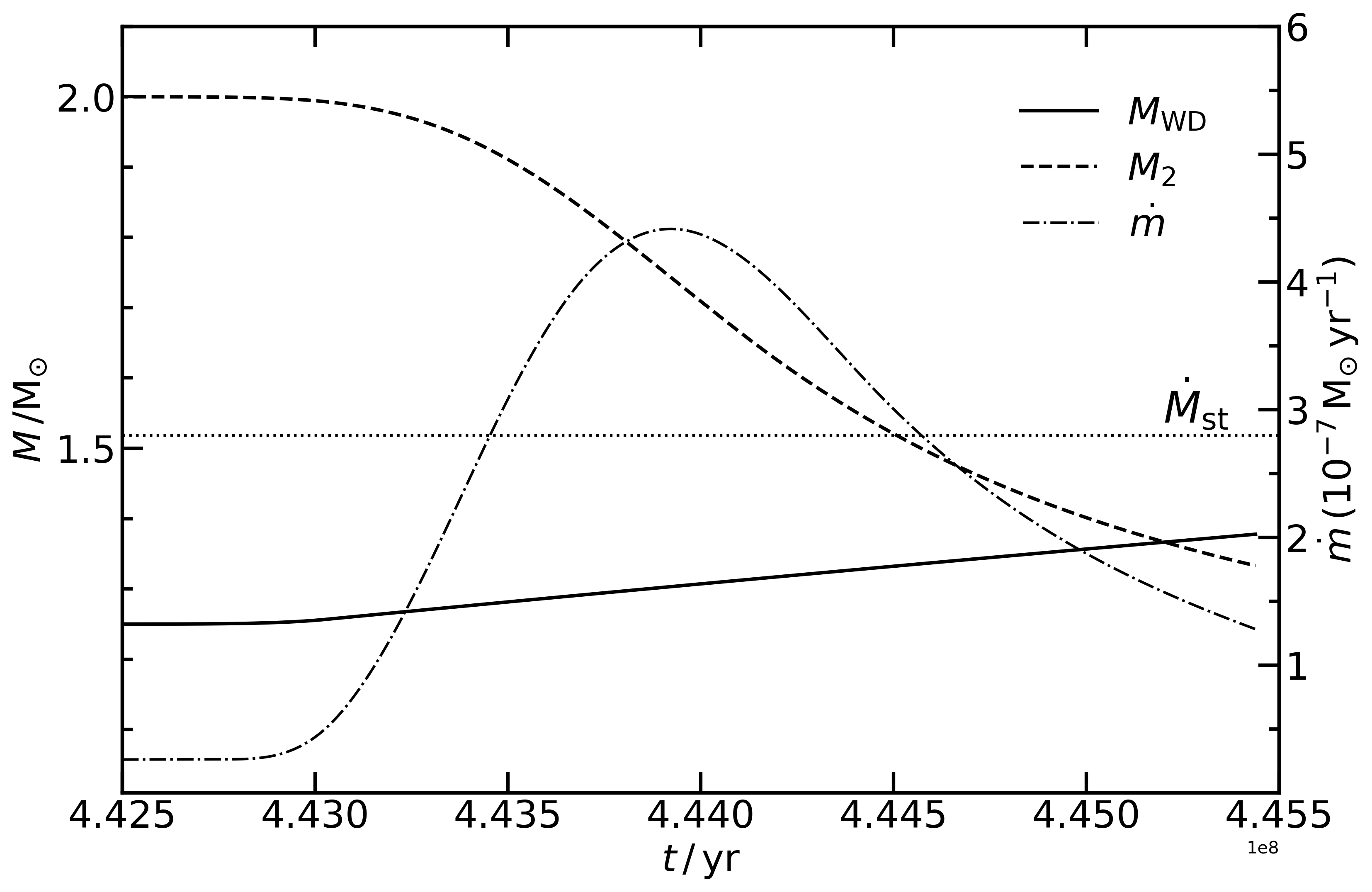}
        \caption{Example of the evolution of a binary. The solid line, dashed line, and dash-dot line represent the mass of the WD ($M_{\rm WD}$), the mass of the companion star ($M_{\rm 2}$), and the mass transfer rate between the binary ($\dot{m}$), respectively. The initial mass of the WD is 1.25 ${\rm M}_{\odot}$, and the mass of the companion star is 2.0 ${\rm M}_{\odot}$. The initial binary orbital period is 1.7 d.  Here, the $\dot{M}_{\rm st}$ represents the stable accretion rate onto the WD.}
        \label{fig:example} 
   \end{figure} 
   
 In our model, the irradiation-driven mass transfer cycle begins with stable hydrogen burning in a WD emitting supersoft X-rays. Stable hydrogen burning on the surface of the WD requires a continuous accretion of hydrogen-rich material from its companion star ($\sim 10^{-7}\,{\rm M}_{\odot}\,{\rm yr}^{-1}$) until a critical mass is accumulated, triggering stable hydrogen burning. This requires that the mass transfer rate reaches a stable accretion rate onto the WD and that it persists for several tens of years ($\sim$ 100 years for $M_{\rm WD} = 1.0\, {\rm M}_{\odot}$). For a detailed understanding of the evolution of the WD and its companion star in an SSS system, we present Fig.~\ref{fig:example}.

 In Fig.~\ref{fig:example}, we show the evolutionary tracks of the WD and its companion star in a binary system. The initial mass of the WD is 1.25 ${\rm M}_{\odot}$, while the companion is a 2.0 ${\rm M}_{\odot}$ evolved MS star that is already out of thermodynamic equilibrium and has lost its MS characteristics. Approximately $4.4 \times 10^8$ years into the MS evolution, as a consequence of the evolution, the companion fills its Roche lobe, initiating thermal timescale mass transfer. As it continues to evolve, it expands and overflows its Roche lobe, leading to an increase in the mass transfer rate. The mass transfer rate can reach the stable accretion rate onto the WD, peaking at around $4.5 \times 10^{-7}\,{\rm M}_{\odot}\,{\rm yr}^{-1}$ and persisting for over $10^{6}$ years. Due to mass loss from the companion and a decrease in its own mass, the mass transfer rate gradually declines, eventually falling below the stable accretion rate onto the WD. During this phase, the WD accretes material from the companion, gradually growing its mass to the Chandrasekhar mass limit.

\subsection{Irradiation-driven mass transfer cycle}
\title{sss}

 Figure~\ref{fig:example} shows the evolution of a binary without the irradiation effects. As the companion star nuclear evolves, it expands to overflow its Roche lobe, leading to the mass transfer rate reaches the stable accretion rate onto the WD for a duration of $10^{6}$ years. Based on this, we added periodic X-ray irradiation on the companion star and examined whether or not the periodic X-ray irradiation can reproduce a periodic mass transfer rate.

\begin{figure}
        \centering
        \includegraphics[width=0.99\columnwidth]{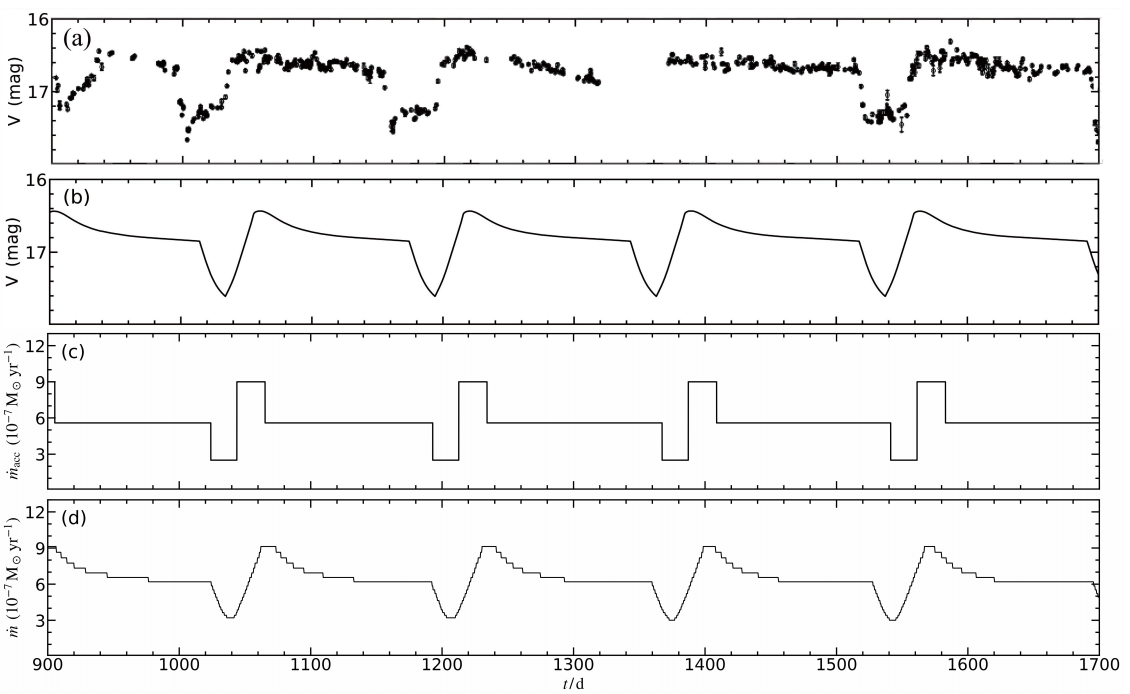}
        \caption{Comparison between the observed V-band light curve of RX~J0513.9-6951 (panel a; \citealt{Alcock1996}) and that predicted from the WD accretion model (Panel b) \citet[]{Zhao2022}. Panel (c) shows the initial mass accretion rate ($\dot{m}_\mathrm { acc}$) adopted in our model. Panel (d) presents the mass transfer rate that we have calculated in this work for the irradiation efficiency is 0.15, where the WD mass is 1.3 ${\rm M}_{\odot}$ and the mass of the companion star is 1.9 ${\rm M}_{\odot}$. }
        \label{fig:SSS}
   \end{figure}

 Figure~\ref{fig:SSS} shows a comparison between this work and the modeling results from the work of \citet[]{Zhao2022}. RX J0513.9-6951 is an SSS that has a 
 periodic optical light curve. \citet[]{Zhao2022} have assumed a periodic WD mass accretion rate (mass transfer rate) and successfully reproduced a periodic optical light curve (panel a). In this work, we calculated the X-ray luminosity based on the data from the work of \citet[]{Zhao2022}. We calculated a periodic X-ray luminosity and used it to irradiate the companion star. The companion star is irradiated periodically by the X-rays from the accreting WD; it expands or contracts periodically, and then the mass transfer rate increases or decreases periodically. The trend of the mass transfer rate from periodic X-ray irradiation is quite similar to that of the accretion rate from the work of \citet[]{Zhao2022}.

 It is important to note that the period of the optical light curve is strongly dependent on the WD mass. The period of the light curve in RX J0513.9-6951 is well determined, about 100 to 190 days \citep{Alcock1996,Cowley2002,Reinsch1996,Reinsch2000}. To fit the observed light curve, the period of the mass transfer cycle must match the observed period of the optical light curve. Due to the periodic variation in mass transfer, the irradiation from the WD also varies periodically. Therefore, the period of the mass transfer and the irradiation are also dependent on the WD mass. In this work, the WD mass in SSS RX J0513.9-6951 is determined to be $1.3\,{\rm M}_{\odot}$ \citep{Zhao2022}. 

 When the mass transfer rate between binary stars continuously exceeds the stable accretion rate onto the WD for $10^{3}$ years, the photosphere of the WD begins stable hydrogen burning. During this time, the surface of the WD emits a substantial amount of supersoft X-rays. These intense supersoft X-rays irradiate the companion star, causing the surface material of the companion star to be heated, and then the companion star expands in a short time (on a timescale significantly shorter than the thermal timescale). This results in an increased overflow of companion material beyond the Roche lobe. Consequently, the mass transfer rate rapidly increases, exceeding the critical accretion rate onto the WD. The WD then burns the surface material at the critical accretion rate onto the WD, and the material that exceeds this critical accretion rate onto the WD accumulates on the surface of the WD. Consequently, the photosphere of the WD gradually expands, increasing its radius and decreasing its effective temperature. The flux of X-rays decreases with the reduction in effective temperature. Due to the decreased X-ray flux, the companion star gradually contracts within its thermal timescale, causing the mass transfer rate to gradually drop below the critical accretion rate onto the WD. During this process, the WD slowly burns the accumulated material on its surface. Once the accumulated material is depleted, the photosphere of the WD rapidly contracts, and it returns to its initial state of emitting a significant amount of supersoft X-rays. A new cycle then begins. We reproduced the periodic mass transfer rate by periodic X-rays heating the companion star, which may be the origin of the periodic accretion rate onto the WD.

\subsection{The irradiation efficiency}
\title{irr_energy}

 \begin{figure}
        \centering
        \includegraphics[width=0.80\columnwidth]{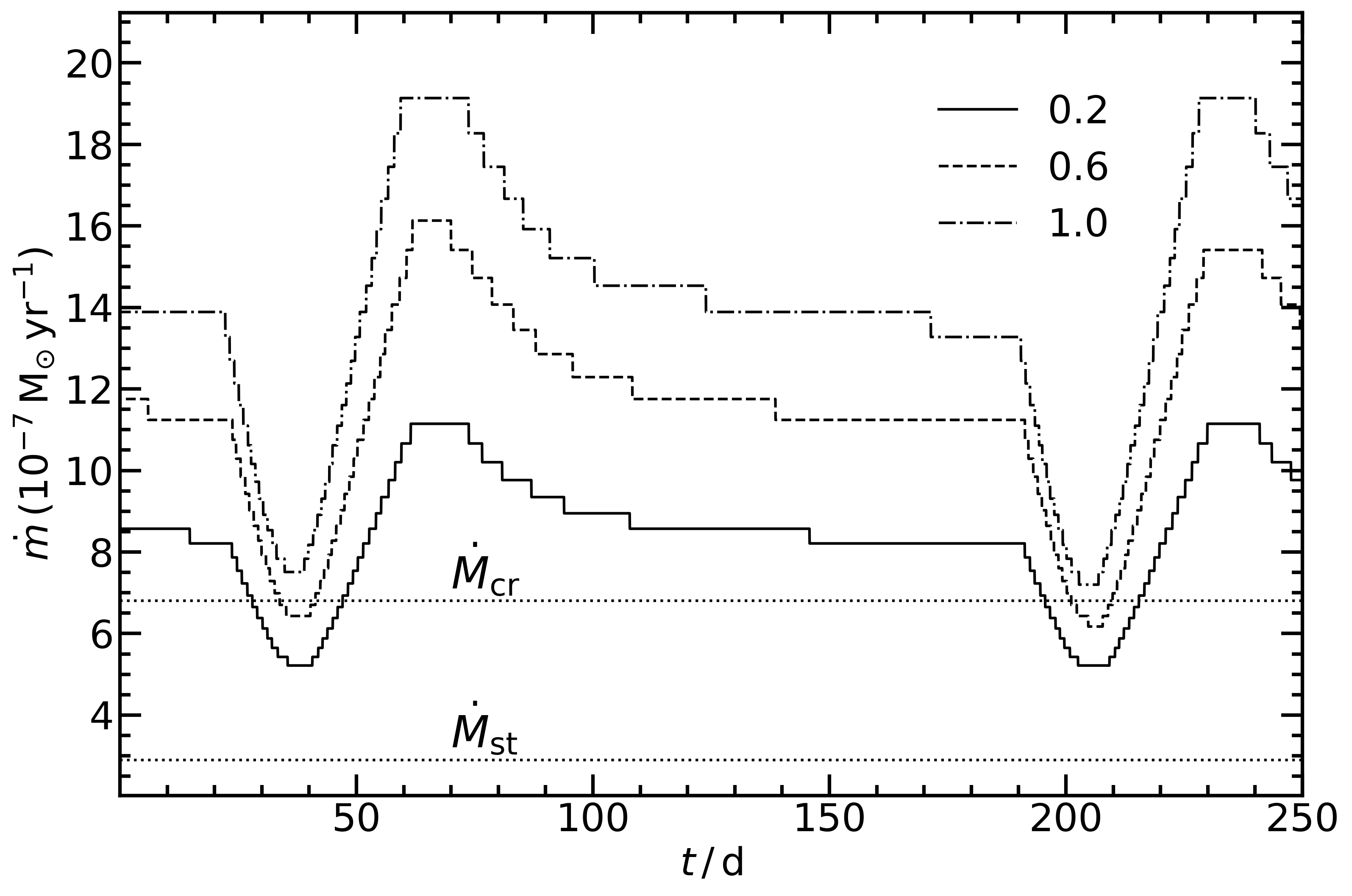}
        \caption{Mass transfer rate cycle on our model for different irradiation efficiencies. Here, the mass of the companion star is 1.9 ${\rm M}_{\odot}$. The solid, dashed, and dotted lines represent irradiation efficiencies of 0.2, 0.6, and 1.0, respectively. Here, the $\dot{M}_{\rm st}$ and $\dot{M}_{\rm cr}$ represent the stable accretion rate and the critical accretion rate onto the WD, respective.}
        \label{fig:flux}
   \end{figure}

 In our model, energy is injected into the surface layer of the companion star, causing the surface layer to expand and overflow its Roche lobe. In their work, \citet[]{Kovetz1988} provided a relation $\dot{m} \propto T_{\rm irr}^{5/3}$ to estimate the equilibrium mass transfer rate $\dot{m}$, where $T_{\rm irr}$ is the outer boundary temperature derived from the irradiation flux $\sim \sigma T_{\rm irr}^{4}$. According to this relation, for a companion with a given companion star mass, the transfer rate increases with a higher irradiation energy (irradiation efficiency). 
 
 When X-rays from the WD irradiate the companion star, the accretion disk partially obscures some of the X-rays, where irradiation efficiency represents the proportion of energy that reaches the surface of the companion star after being blocked by the accretion disk of the WD. In our work, we examined the impact of different irradiation efficiencies on the mass transfer rate. In Fig.~\ref{fig:flux}, the mass transfer rate is plotted against the irradiation energy for a fixed companion mass of 1.9 ${\rm M}_{\odot}$. The initial mass transfer rate is $3.0 \times 10^{-7}\,{\rm M}_{\odot}\,{\rm yr}^{-1}$ (the stable accretion rate onto the WD with 1.3 ${\rm M}_{\odot}$). The solid, dashed, and dotted lines represent irradiation efficiencies of 0.2, 0.6, and 1.0 respectively. According to Fig.~\ref{fig:flux}, the mass transfer rate increases with the irradiation efficiency. However, the irradiation efficiency does not significantly affect the mass transfer cycle. The increase in irradiation efficiency implies that more irradiation energy reaches the surface of the companion star, leading to increased surface energy of the companion star. The companion star expands, causing more material to overflow the Roche lobe and thereby increasing the mass transfer rate.

 The radius of an NS in low-mass X-ray binaries is typically very small, usually only around $\sim 10 \,{\rm km}$. Due to this small size, most of the X-rays emitted by the NS are blocked by its accretion disk, and only a small portion can reach the surface of the companion star. Previous studies have suggested the irradiation efficiencies may be less than  $\sim 10{\%}$ \citep{Harpaz1994}. In contrast, for the WDs in SSSs, their radii are much larger than those of NSs (typically $\sim 0.01 \,R_{\odot}$). As a result, most of the supersoft X-rays may not be blocked by the accretion disk of the WD. Therefore, the irradiation efficiency is assumed to be lower than 1.0 here.

 This result implies that for a given companion star mass, the higher irradiation energies can result in higher mass transfer rates. For an SSS with a WD of 1.3 ${\rm M}_{\odot}$ and a companion star of 1.9 ${\rm M}_{\odot}$, the irradiation efficiency may be 0.15. Due to the periodic X-ray flux in SSSs, this mechanism could provide an explanation for the origin of periodic mass transfer rates. In other words, periodic X-ray irradiation on the companion star can induce a periodic mass transfer rate, that is, the mass transfer rate cycle in Fig.~\ref{fig:SSS}.

\subsection{The companion mass}

 \begin{figure}
        \centering
        \includegraphics[width=0.80\columnwidth]{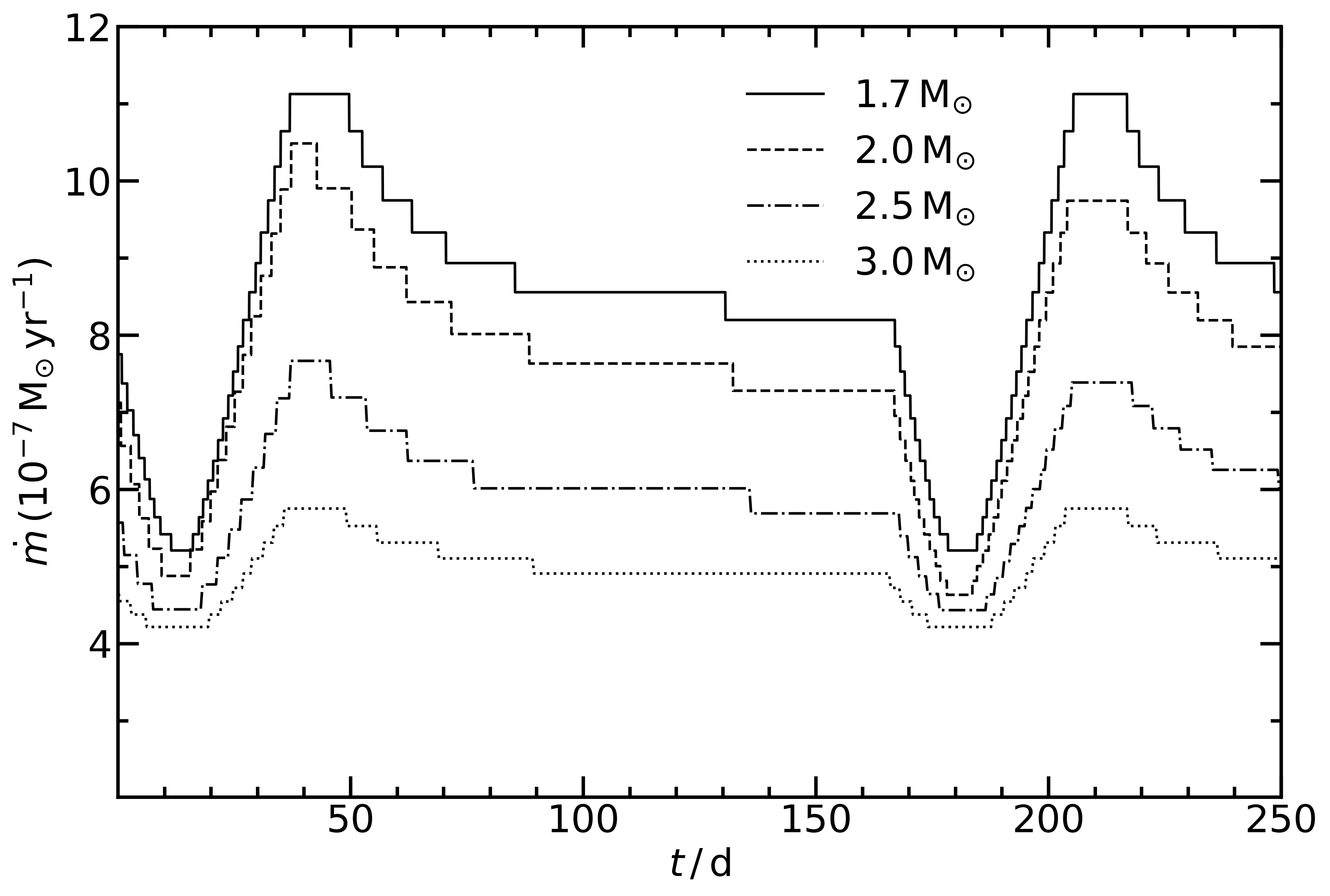}
        \caption{Mass transfer rate cycle on our model for different companion masses for a given irradiation efficiency of 0.2. The solid, dashed, dash-dotted, and dotted lines represent a companion mass of 1.7, 2.0, 2.5, and 3.0 ${\rm M}_{\odot}$, respectively.}
        \label{fig:com}
   \end{figure} 
   
 In binary star systems, the mass transfer rate between the component stars is mainly determined by the mass ratio, the masses of the stars, and the orbital period. Therefore, for a system with a given WD mass and a given orbital period, the mass of the companion star significantly influences the mass transfer rate. To understand the relationship between the mass of the companion star and the equilibrium mass transfer rate, we refer to Fig.~\ref{fig:com}. This figure illustrates the variation of the mass transfer cycle with the mass of the companion star for a given irradiation efficiency of 0.2.

 Previous works have suggested that the companion star mass in an SSS falls between 1.5 and 2.5 ${\rm M}_{\odot}$ \citep{PH.2010}. 
 In our work, the initial model of a WD steadily burning hydrogen requires the mass transfer rate between the binary stars to reach the stable accretion rate onto the WD for a certain duration. For a WD with 1.3 ${\rm M}_{\odot}$, if the mass of the companion star is below 1.7 ${\rm M}_{\odot}$, the mass transfer rate will fall below the stable accretion rate into the WD. \citet[]{Hachisu2003c, Hachisu2003b} provide models with a companion star mass of 2.6 ${\rm M}_{\odot}$. In our model, we considered a slightly larger range of companion star masses, from 1.7 to 3.0 ${\rm M}_{\odot}$. Based on Fig.~\ref{fig:com}, we find that the mass transfer rate becomes lower with the companion star mass for a given irradiation efficiency. The greater the mass of the companion star, the less pronounced the variations in the mass transfer cycle. The relationship is thus toward a linear trend.
 
 Regarding the case of a more massive companion star, its nuclear reactions are more intense compared to those of a low-mass companion star, resulting in higher effective temperatures. For the same X-ray flux, the higher the effective temperature of the companion star, the weaker the heating effect of the X-rays. It is evident that if the mass of the companion star is sufficiently large (e.g., 4.0 ${\rm M}_{\odot}$), the surface flux far exceeds the X-ray irradiation flux, resulting in the heating effect of X-rays being very weak. In our model, we suggest that in SSSs, lower-mass companion stars are more likely to produce significant mass transfer cycles due to the irradiation effect of X-rays. For example, the mass of the companion star in SSS RX J0513.9-6951 is likely to be less than 2.5 ${\rm M}_{\odot}$.

\section{Discussion}
\label{discussion}

  \citet{Zhao2022} reproduced the quasi-periodic optical light curve in SSSs by assuming a periodic mass transfer. In this work, we have investigated the impact of X-ray irradiation on companion stars regarding their mass transfer process. We find that periodic X-ray irradiation of the companion star can induce periodic expansion or contraction of the companion star, thereby generating mass transfer cycles. For a given companion star, higher irradiation efficiency results in a higher mass transfer rate. Additionally, the mass transfer rate increases as the mass of the companion star decreases for a given irradiation efficiency. We propose that mass transfer driven by periodic X-ray irradiation could be the origin of the quasi-periodic optical light curve observed in SSSs. However, in our model, we treated the WD as a point mass and only focused on the detailed evolution of the donor star. Therefore, we could not calculate a complete cycle that includes the detailed evolution of both the WD and the donor star. Future research should aim to develop more comprehensive models that fully integrate the simultaneous evolution of both stars.

  In the context of SSSs, previous studies have suggested that the companion star should be more massive than 1.5\,${\rm M}_{\odot}$ \citep[]{vandenHeuvel1992, vanTeeseling1996, PH.2010}. However, considering the irradiation of the companion star by X-rays, the lower-mass companion star has a more pronounced irradiation effect than a more massive companion star for a given WD. The mass transfer rate between the binary increases as the companion star mass decreases. This implies that even the companion star mass is lower than 1.5 \,${\rm M}_{\odot}$. If the companion star is irradiated by a high flux, it is still capable of providing a sufficiently high mass transfer rate. Such a mass transfer rate allows the WD to stably burn hydrogen on its surface. For instance, the companion star may be a low-mass MS star ($< 1.0 \,{\rm M}_{\odot}$) in a nova system. The WD exhibits high luminosity close to the Eddington luminosity during the nova eruption. The companion star may experience more noticeable irradiation effects under such conditions, leading to a higher mass transfer rate, possibly exceeding the stable hydrogen burn rate onto the WD \citep{Ginzburg2021}.

  People have estimated the contribution of the SD model of SNe Ia by counting the number of SSSs. The actual observed number of SSSs is lower than what is needed to explain the observed rate of SNe Ia \citep{Gilfanov2010Nature}. However, some studies suggest that this might overestimate the time that mass-accretion WDs spend as SSSs \citep{Meng2010,Hachisu2010}. Our results show that the mass transfer rate exceeds the critical accretion rate onto the WD most of the time, which means the time spent in the SSS phase is extremely short. Therefore, the number of SSSs that can be observed and identified is much less than previously estimated.

\section{Summary}
\label{summary}
  In an SSS system comprising a WD and a massive companion star, the companion star is irradiated and heated by the supersoft X-rays emitted by the WD. Our work demonstrates that these X-rays can alter the surface boundary condition of the companion star, causing it to expand and overflow the Roche lobe and then increase the mass transfer rate. We also find that for a given companion star, a higher irradiation energy results in a higher mass transfer rate. For a given irradiation energy, the mass transfer rate increases with the companion star. This suggests that periodic X-ray irradiation can result in corresponding periodic changes in mass transfer. We have reproduced mass transfer cycles by simulating the feedback of the companion stars under periodic X-ray irradiation, providing an explanation for the variable accretion rates onto the WDs in SSSs. This suggests that mass transfer driven by X-ray irradiation onto the companion star may be the origin of the quasi-periodic optical light curve observed in SSSs.

\begin{acknowledgements}
    We acknowledge the anonymous referee for the valuable comments that helped improve this paper. This work is supported by the National Natural Science Foundation of China (Nos. 12333008 and 12288102) and National Key R\&D Program of China (No. 2021YFA1600403). X.M. acknowledges support from the CAS “Light of West China” Program. X.M. acknowledges support from Yunnan Fundamental Research Projects (Nos. 202401BC070007 and 202201BC070003), International Centre of Supernovae, Yunnan Key Laboratory (No. 202302AN360001), the Yunnan Revitalization Talent Support Program Science \& Technology Champion Project (NO. 202305AB350003), and the science research grants from the China Manned Space Project.

\end{acknowledgements} 
\bibliographystyle{aa}
\bibliography{References} 

\begin{thebibliography}{47}
\expandafter\ifx\csname natexlab\endcsname\relax\def\natexlab#1{#1}\fi

\bibitem[{{Alcock} {et~al.}(1996){Alcock}, {Allsman}, {Alves}, {Axelrod},
  {Bennett}, {Charles}, {Cook}, {Freeman}, {Griest}, {Guern}, {Lehner},
  {Livio}, {Marshall}, {Peterson}, {Pratt}, {Quinn}, {Rodgers}, {Southwell},
  {Stubbs}, {Sutherland}, \& {Welch}}]{Alcock1996}
{Alcock}, C., {Allsman}, R.~A., {Alves}, D., {et~al.} 1996, \mnras, 280, L49

\bibitem[{{Branch} {et~al.}(1995){Branch}, {Livio}, {Yungelson}, {Boffi}, \&
  {Baron}}]{Branch1995}
{Branch}, D., {Livio}, M., {Yungelson}, L.~R., {Boffi}, F.~R., \& {Baron}, E.
  1995, \pasp, 107, 1019

\bibitem[{{Cowley} {et~al.}(2002){Cowley}, {Schmidtke}, {Crampton}, \&
  {Hutchings}}]{Cowley2002}
{Cowley}, A.~P., {Schmidtke}, P.~C., {Crampton}, D., \& {Hutchings}, J.~B.
  2002, \aj, 124, 2233

\bibitem[{{G{\"a}nsicke} {et~al.}(1998){G{\"a}nsicke}, {van Teeseling},
  {Beuermann}, \& {de Martino}}]{Gansicke1998}
{G{\"a}nsicke}, B.~T., {van Teeseling}, A., {Beuermann}, K., \& {de Martino},
  D. 1998, \aap, 333, 163

\bibitem[{{Gilfanov} \& {Bogd{\'a}n}(2010)}]{Gilfanov2010Nature}
{Gilfanov}, M. \& {Bogd{\'a}n}, {\'A}. 2010, \nat, 463, 924

\bibitem[{{Ginzburg} \& {Quataert}(2021)}]{Ginzburg2021}
{Ginzburg}, S. \& {Quataert}, E. 2021, \mnras, 507, 475

\bibitem[{{Greiner} {et~al.}(1991){Greiner}, {Hasinger}, \&
  {Kahabka}}]{Greiner1991}
{Greiner}, J., {Hasinger}, G., \& {Kahabka}, P. 1991, \aap, 246, L17

\bibitem[{{Hachisu} \& {Kato}(2003{\natexlab{a}})}]{Hachisu2003c}
{Hachisu}, I. \& {Kato}, M. 2003{\natexlab{a}}, \apj, 598, 527

\bibitem[{{Hachisu} \& {Kato}(2003{\natexlab{b}})}]{Hachisu2003b}
{Hachisu}, I. \& {Kato}, M. 2003{\natexlab{b}}, \apj, 590, 445

\bibitem[{{Hachisu} {et~al.}(2010){Hachisu}, {Kato}, \& {Nomoto}}]{Hachisu2010}
{Hachisu}, I., {Kato}, M., \& {Nomoto}, K. 2010, \apjl, 724, L212

\bibitem[{{Hameury} {et~al.}(1993){Hameury}, {King}, {Lasota}, \&
  {Raison}}]{Hameury1993}
{Hameury}, J.~M., {King}, A.~R., {Lasota}, J.~P., \& {Raison}, F. 1993, \aap,
  277, 81

\bibitem[{{Hameury} \& {Ritter}(1997)}]{Hameury1997}
{Hameury}, J.~M. \& {Ritter}, H. 1997, \aaps, 123, 273

\bibitem[{{Harpaz} \& {Rappaport}(1991)}]{Harpaz1991}
{Harpaz}, A. \& {Rappaport}, S. 1991, \apj, 383, 739

\bibitem[{{Harpaz} \& {Rappaport}(1994)}]{Harpaz1994}
{Harpaz}, A. \& {Rappaport}, S. 1994, \apj, 434, 283

\bibitem[{{Hillebrandt} \& {Niemeyer}(2000)}]{Hillebrandt2000}
{Hillebrandt}, W. \& {Niemeyer}, J.~C. 2000, \araa, 38, 191

\bibitem[{{Howell}(2011)}]{Howell2011}
{Howell}, D.~A. 2011, Nature Communications, 2, 350

\bibitem[{{Kahabka} \& {van den Heuvel}(1997)}]{Kahabka1997}
{Kahabka}, P. \& {van den Heuvel}, E.~P.~J. 1997, \araa, 35, 69

\bibitem[{{King} {et~al.}(1995){King}, {Frank}, {Kolb}, \& {Ritter}}]{King1995}
{King}, A.~R., {Frank}, J., {Kolb}, U., \& {Ritter}, H. 1995, \apjl, 444, L37

\bibitem[{{Kolb} \& {Ritter}(1990)}]{Kolb1990}
{Kolb}, U. \& {Ritter}, H. 1990, \aap, 236, 385

\bibitem[{{Kovetz} {et~al.}(1988){Kovetz}, {Prialnik}, \& {Shara}}]{Kovetz1988}
{Kovetz}, A., {Prialnik}, D., \& {Shara}, M.~M. 1988, \apj, 325, 828

\bibitem[{{Liu} \& {Stancliffe}(2018)}]{Liu2018}
{Liu}, Z.-W. \& {Stancliffe}, R.~J. 2018, \mnras, 475, 5257

\bibitem[{{Maoz} {et~al.}(2014){Maoz}, {Mannucci}, \& {Nelemans}}]{Maoz2014}
{Maoz}, D., {Mannucci}, F., \& {Nelemans}, G. 2014, \araa, 52, 107

\bibitem[{{Mason}(1989)}]{Mason1989}
{Mason}, K.~O. 1989, in ESA Special Publication, Vol.~1, Two Topics in X-Ray
  Astronomy, Volume 1: X Ray Binaries. Volume 2: AGN and the X Ray Background,
  ed. J.~{Hunt} \& B.~{Battrick}, 113

\bibitem[{{Meng} {et~al.}(2009){Meng}, {Chen}, \& {Han}}]{Meng2009a}
{Meng}, X., {Chen}, X., \& {Han}, Z. 2009, \mnras, 395, 2103

\bibitem[{{Meng} \& {Yang}(2010)}]{Meng2010}
{Meng}, X. \& {Yang}, W. 2010, \apj, 710, 1310

\bibitem[{{Meyer-Hofmeister} {et~al.}(1997){Meyer-Hofmeister}, {Schandl}, \&
  {Meyer}}]{Meyer-Hofmeister1997}
{Meyer-Hofmeister}, E., {Schandl}, S., \& {Meyer}, F. 1997, \aap, 321, 245

\bibitem[{{Milgrom}(1978)}]{Milgrom1978}
{Milgrom}, M. 1978, \aap, 67, L25

\bibitem[{{Nomoto} {et~al.}(2007){Nomoto}, {Saio}, {Kato}, \&
  {Hachisu}}]{Nomoto2007}
{Nomoto}, K., {Saio}, H., {Kato}, M., \& {Hachisu}, I. 2007, \apj, 663, 1269

\bibitem[{{Nomoto} {et~al.}(1984){Nomoto}, {Thielemann}, \&
  {Yokoi}}]{Nomoto1984}
{Nomoto}, K., {Thielemann}, F.~K., \& {Yokoi}, K. 1984, \apj, 286, 644

\bibitem[{{Pakull} {et~al.}(1993){Pakull}, {Motch}, {Bianchi}, {Thomas},
  {Guibert}, {Beaulieu}, {Grison}, \& {Schaeidt}}]{Pakull1993}
{Pakull}, M.~W., {Motch}, C., {Bianchi}, L., {et~al.} 1993, \aap, 278, L39

\bibitem[{{Paxton} {et~al.}(2011){Paxton}, {Bildsten}, {Dotter}, {Herwig},
  {Lesaffre}, \& {Timmes}}]{Paxton2011}
{Paxton}, B., {Bildsten}, L., {Dotter}, A., {et~al.} 2011, \apjs, 192, 3

\bibitem[{{Paxton} {et~al.}(2013){Paxton}, {Cantiello}, {Arras}, {Bildsten},
  {Brown}, {Dotter}, {Mankovich}, {Montgomery}, {Stello}, {Timmes}, \&
  {Townsend}}]{Paxton2013}
{Paxton}, B., {Cantiello}, M., {Arras}, P., {et~al.} 2013, \apjs, 208, 4

\bibitem[{{Paxton} {et~al.}(2015){Paxton}, {Marchant}, {Schwab}, {Bauer},
  {Bildsten}, {Cantiello}, {Dessart}, {Farmer}, {Hu}, {Langer}, {Townsend},
  {Townsley}, \& {Timmes}}]{Paxton2015}
{Paxton}, B., {Marchant}, P., {Schwab}, J., {et~al.} 2015, \apjs, 220, 15

\bibitem[{{Paxton} {et~al.}(2018){Paxton}, {Schwab}, {Bauer}, {Bildsten},
  {Blinnikov}, {Duffell}, {Farmer}, {Goldberg}, {Marchant}, {Sorokina},
  {Thoul}, {Townsend}, \& {Timmes}}]{Paxton2018}
{Paxton}, B., {Schwab}, J., {Bauer}, E.~B., {et~al.} 2018, \apjs, 234, 34

\bibitem[{{Paxton} {et~al.}(2019){Paxton}, {Smolec}, {Schwab}, {Gautschy},
  {Bildsten}, {Cantiello}, {Dotter}, {Farmer}, {Goldberg}, {Jermyn}, {Kanbur},
  {Marchant}, {Thoul}, {Townsend}, {Wolf}, {Zhang}, \& {Timmes}}]{Paxton2019}
{Paxton}, B., {Smolec}, R., {Schwab}, J., {et~al.} 2019, \apjs, 243, 10

\bibitem[{{Perlmutter} {et~al.}(1999){Perlmutter}, {Aldering}, {Goldhaber},
  {Knop}, {Nugent}, {Castro}, {Deustua}, {Fabbro}, {Goobar}, {Groom}, {Hook},
  {Kim}, {Kim}, {Lee}, {Nunes}, {Pain}, {Pennypacker}, {Quimby}, {Lidman},
  {Ellis}, {Irwin}, {McMahon}, {Ruiz-Lapuente}, {Walton}, {Schaefer}, {Boyle},
  {Filippenko}, {Matheson}, {Fruchter}, {Panagia}, {Newberg}, {Couch}, \&
  {Project}}]{Perlmutter1999}
{Perlmutter}, S., {Aldering}, G., {Goldhaber}, G., {et~al.} 1999, \apj, 517,
  565

\bibitem[{{Podsiadlowski}(1991)}]{PH.1991}
{Podsiadlowski}, P. 1991, \nat, 350, 136

\bibitem[{{Podsiadlowski}(2010)}]{PH.2010}
{Podsiadlowski}, P. 2010, Astronomische Nachrichten, 331, 218

\bibitem[{{Reinsch} {et~al.}(1996){Reinsch}, {van Teeseling}, {Beuermann}, \&
  {Abbott}}]{Reinsch1996}
{Reinsch}, K., {van Teeseling}, A., {Beuermann}, K., \& {Abbott}, T.~M.~C.
  1996, \aap, 309, L11

\bibitem[{{Reinsch} {et~al.}(2000){Reinsch}, {van Teeseling}, {King}, \&
  {Beuermann}}]{Reinsch2000}
{Reinsch}, K., {van Teeseling}, A., {King}, A.~R., \& {Beuermann}, K. 2000,
  \aap, 354, L37

\bibitem[{{Riess} {et~al.}(1998){Riess}, {Filippenko}, {Challis},
  {Clocchiatti}, {Diercks}, {Garnavich}, {Gilliland}, {Hogan}, {Jha},
  {Kirshner}, {Leibundgut}, {Phillips}, {Reiss}, {Schmidt}, {Schommer},
  {Smith}, {Spyromilio}, {Stubbs}, {Suntzeff}, \& {Tonry}}]{Riess1998}
{Riess}, A.~G., {Filippenko}, A.~V., {Challis}, P., {et~al.} 1998, \aj, 116,
  1009

\bibitem[{{Ritter} {et~al.}(2000){Ritter}, {Zhang}, \& {Kolb}}]{Ritter2000}
{Ritter}, H., {Zhang}, Z.~Y., \& {Kolb}, U. 2000, \aap, 360, 959

\bibitem[{{Southwell} {et~al.}(1996){Southwell}, {Livio}, {Charles},
  {O'Donoghue}, \& {Sutherland}}]{Southwell1996}
{Southwell}, K.~A., {Livio}, M., {Charles}, P.~A., {O'Donoghue}, D., \&
  {Sutherland}, W.~J. 1996, \apj, 470, 1065

\bibitem[{{van den Heuvel} {et~al.}(1992){van den Heuvel}, {Bhattacharya},
  {Nomoto}, \& {Rappaport}}]{vandenHeuvel1992}
{van den Heuvel}, E.~P.~J., {Bhattacharya}, D., {Nomoto}, K., \& {Rappaport},
  S.~A. 1992, \aap, 262, 97

\bibitem[{{van Teeseling} {et~al.}(1996){van Teeseling}, {Beuermann}, \&
  {Verbunt}}]{vanTeeseling1996}
{van Teeseling}, A., {Beuermann}, K., \& {Verbunt}, F. 1996, \aap, 315, 467

\bibitem[{{Whelan} \& {Iben}(1973)}]{Whelan1973}
{Whelan}, J. \& {Iben}, Icko, J. 1973, \apj, 186, 1007

\bibitem[{{Zhao} {et~al.}(2022){Zhao}, {Meng}, {Cui}, \& {Liu}}]{Zhao2022}
{Zhao}, W., {Meng}, X., {Cui}, Y., \& {Liu}, Z.-W. 2022, \aap, 666, A81

\end{thebibliography}
\end{document}